# Enhanced ammonia electro-oxidation reaction on platinum-iron oxide catalyst assisted by MagnetoElectroCatalysis

Caio Machado Fernandes[a,d*], Eduardo M. Rodrigues[a], Odivaldo C. Alves[a], Flavio Garcia[b], Yutao Xing[c], Mauro C. dos Santos[d], Júlio César M. Silva[a]

[a]*Department of Physical Chemistry, Fluminense Federal University, Campus Valonguinho, 24020-141, Niterói, RJ, Brazil.*
[b]*Brazilian Center for Research in Physics, Urca, 22290-180, Rio de Janeiro, RJ, Brazil.*
[c]*High-resolution Electron Microscopy Laboratory, Advanced Characterization Center for Petroleum Industry, Fluminense Federal University, 24210-346, Niterói, RJ, Brazil*
[d]*Laboratory of Electrochemistry and Nanostructured Materials, Center for Natural and Human Sciences, Federal University of ABC, 09210-170, Santo André, SP, Brazil*

*Corresponding author e-mail: c.fernandes@ufabc.edu.br

**Abstract**

Ammonia poses significant environmental challenges due to its role in water pollution, contributing to eutrophication and several detrimental environmental and ecological issues. Addressing the efficient removal or conversion of ammonia is, therefore, critical. Among various methods, the ammonia electro-oxidation reaction stands out due to its potential for direct energy conversion and environment remediation. Here, we synthesize platinum-iron oxide magnetic nanoparticles (Pt-MNP) as electrocatalysts and apply an alternating magnetic field (AMF) to enhance their activity.. The AMF generates localized heat via Néel relaxation, accelerating ammonia oxidation kinetics at the catalytic surface.. Compared to conventional electro-oxidation methods, this technique demonstrates superior efficiency and stability, offering a promising alternative for ammonia treatment. This work uses the concept of MagnetoElectroCatalysis, showcasing the synergy between magnetic fields and the electrochemical process, leveraging the AMF to induce localized heating within the nanocatalyst, thereby improving its catalytic activity as shown in cyclic voltammetry and chronoamperometry experiments. By combining nanocatalyst design with innovative AMF application, this study provides a new avenue for enhancing electrochemical reactions, with broad implications for environmental remediation and sustainable energy solutions.

**Keywords:** platinum; platinum-iron oxide nanoparticles; electrocatalysts; ammonia electro-oxidation reaction; magnetic hyperthermia; MagnetoElectroCatalysis.

# 1. Introduction

The global nitrogen cycle is facing a severe imbalance, escalating rapidly due to the uncontrolled discharge of nitrogenous pollutants. Among these, ammonia-nitrogen ($NH_4^+ - N$) has emerged as a major contributor primarily from sources such as domestic sewage, industrial wastewater, and agricultural runoff [1]. The detrimental effects of ammonia on aquatic ecosystems cannot be ignored, as it significantly promotes eutrophication and disrupts dissolved oxygen levels, posing a grave threat to aquatic species [2]. The urgent need to address this challenge has spurred considerable attention towards removing ammonia from diverse wastewaters. The amplification of ammonia concentrations in natural waters severely threatens ecological equilibrium, underscoring the urgent need to tackle this issue effectively [3].

Various conventional technologies, such as biological treatment, breakpoint chlorination, and air striping, have been relied upon in wastewater treatment [4-6]. However, the emergence of electrochemical oxidation technology presents a compelling and promising alternative. This innovative method offers multiple advantages, including its simplicity in operation, cost-effectiveness, and remarkable tolerance to toxic pollutants. As a result, it has swiftly risen as a frontrunner in the pursuit of effective solutions to combat pollution challenges [7-9].

Among the environmental concerns that electrochemical oxidation addresses with finesse is ammonia pollution. This cutting-edge approach harnesses the power of electrochemical reactions to tackle the ever-growing issue of ammonia-induced ecological imbalances in natural waters. By adopting this sustainable and environmentally friendly solution, we can take a decisive step towards restoring equilibrium to our delicate ecosystems and preserving the health of our planet for generations to come [10-12].

Direct electro-oxidation is an environmentally friendly and cost-effective method involving the catalytic oxidation of ammonia to $N_2$ and $H_2$ directly. This innovative approach bypasses the need for intermediate steps and converts ammonia into harmless nitrogen gas with the intervention of a catalyst [13, 14]. The reaction for direct electrochemical oxidation of ammonia is as follows:

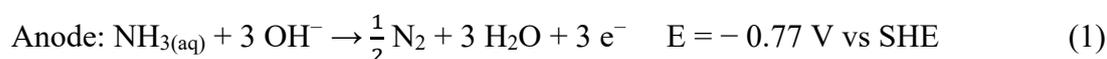

Anode: $NH_{3(aq)} + 3\ OH^- \rightarrow \frac{1}{2} N_2 + 3\ H_2O + 3\ e^-$   $E = -0.77$ V vs SHE   (1)

Cathode: $3\ H_2O + 3\ e^- \rightarrow \frac{3}{2} H_2 + 3\ OH^-$    E = − 0.83 V vs SHE        (2)

Overall reaction: $NH_{3(aq)} \rightarrow \frac{1}{2} N_2 + 3\ H_2$    ΔE = − 0.06 V        (3)

By adopting this effective and sustainable process, we can combat ammonia-related issues effectively, ensuring a cleaner and healthier environment. Moreover, as shown above, the conversion of ammonia into hydrogen and nitrogen gas requires only an external energy of 0.06 V, which is remarkably lower (over 95% reduction) compared to the energy demand for water electrolysis (1.23 V) [15-17]. This significant energy efficiency makes direct electrochemical oxidation an attractive and viable option for environmentally sustainable processes.

Ammonia also holds immense promise as a potential fuel for direct fuel cells, presenting a compelling alternative to conventional hydrogen-based systems. It boasts a hydrogen content of 17.6 wt%, distinguishing it from ethanol and methanol as it does not release $CO_2$ during decomposition [18, 19]. Its impressive 3000 Wh/kg energy density further adds to its allure. In terms of oxidation, ammonia showcases a theoretical charge of 4.75 Ah $g^{-1}$, offering promising possibilities for use in various applications [20]. Additionally, the specific volume of hydrogen in ammonia surpasses liquid hydrogen by a remarkable 70% [21]. These exceptional properties make ammonia a compelling candidate for future energy systems and highlight its potential as a sustainable fuel option.

Platinum is the most suitable catalyst for the ammonia electro-oxidation reaction (AEOR) due to its high catalytic activity and selectivity in producing $N_2$ gas. Typically, Pt is utilized as carbon-supported nanoparticles [22-26]. However, its effectiveness diminishes over time due to $N_{ads}$ poisoning, a byproduct of the AEOR process [27]. Traditionally, the rapid catalytic deactivation of Pt in the AEOR has been attributed to the irreversible adsorption of fully dehydrogenated atomic N, as described by the Gerischer-Mauerer mechanism [28]. More recently, researchers have reported the activity of Pt decay caused by $NO_x$ adsorbates, shedding new light on its deactivation mechanism [29]. Understanding these deactivation pathways is crucial for devising strategies to enhance Pt-based catalysts' long-term stability and efficiency for the ammonia oxidation reaction.

Despite significant efforts invested in defect engineering, phase transition, and doping, progress in electrocatalytic reactions has encountered limitations, primarily dependent on catalyst design and alteration [30, 31]. Furthermore, the advantages of raising the temperature in an electrochemical system are evident for various catalysts and molecules [32-39]. As a specific example, in the study by Zhou and Cheng [40], raising the solution temperature from 22 to 50 °C led to a notable 2.4-fold increase in the maximum current density for ammonia electro-oxidation on Pt catalysts. However, it is essential to consider the substantial energy consumption required to heat the catalytic bed, as this can pose a significant limitation to the overall process efficiency. With that in mind, research groups have been challenged in the past years to overcome those challenges.

Recently, magnetic field-enhanced electrocatalysis has emerged as a novel and promising strategy to significantly improve electrochemical reactions, offering unique advantages over conventional methods [41-46]. Unlike traditional thermal approaches that require bulk heating of the entire catalytic system, the application of an Alternating Magnetic Field (AMF) enables highly localized and energy-efficient heating precisely at the reaction sites. This innovative technique not only minimizes energy waste but also enhances reaction kinetics and selectivity—a critical advancement for processes like ammonia electro-oxidation, where precise thermal control is essential yet challenging to achieve. To the best of our knowledge, the use of AMF for enhancing ammonia electro-oxidation remains underexplored, representing a significant departure from existing methods. By leveraging AMF's ability to selectively activate catalysts without overheating the system, this approach explore new possibilities for optimizing electrochemical ammonia conversion, addressing key limitations in current technologies [47].

Superparamagnetic iron oxide nanoparticles have emerged as promising candidates for magnetic hyperthermia applications. When exposed to an alternating magnetic field (AMF), their magnetic moments rapidly and continuously realign with the external field. This continual realignment induces energy dissipation through Néel (due to the magnetic response to the AMF) and Brownian (due to the friction caused by the movement of MNPs driven by the AMF) relaxation mechanisms, generating highly localized heat. This phenomenon has been extensively explored in biomedical applications, particularly for cancer therapy, but its potential in electrocatalysis remains largely untapped. [48].

By harnessing the power of AMF, remarkable efficiency and unprecedented control over catalytic processes can be achieved, paving the way for more sustainable and impactful advancements. The AMF not only has the capability to induce localized heating in magnetic nanoparticles, thereby increasing the temperature precisely where it is needed, but also facilitates rapid cooling of these nanoparticles once the field is removed. This dual capability is a significant advantage over traditional methods that require heating the entire catalytic bed, which incurs a much higher energy cost. Localized heating ensures that energy is used efficiently, targeting only the active sites of the catalyst. Equally important is the cooling rate, which is just as critical as heating for maintaining optimal reaction conditions and preventing thermal degradation. Unlike conventional systems where cooling the entire bed is necessary, localized heating allows for equally rapid cooling of the nanoparticles, minimizing energy waste and enhancing the overall efficiency of the catalytic process [48]. Indeed, our research group recently published articles using MagnetoElectroCatalysis to boost urea electro-oxidation reaction on nickel-iron oxide catalysts with different morphologies [49, 50].

Iron oxide nanoparticles have emerged as a promising candidate for magnetic hyperthermia [51]. When exposed to an alternating magnetic field (AMF), these nanoparticles exhibit a phenomenon known as superparamagnetism, wherein their magnetic moments repeatedly align and re-align with the magnetic field. This rapid and repeated realignment generates heat through friction and energy dissipation, locally raising the temperature [52-54]. The ability of these nanoparticles to produce controlled heat under AMF excitation holds great potential to be combined with electrochemical experiments and advance this research field. Thus, synthesizing Platinum-Magnetic Nanoparticles (Pt-MNP) material holds immense promise as a groundbreaking strategy for a revolutionary approach to AEOR. Platinum nanoparticles are renowned for their exceptional electrocatalytic activity in driving ammonia oxidation, ensuring efficient and selective conversion to $N_2$ and $H_2$. On the other hand, Magnetic Nanoparticles (MNPs) possess the unique capability of inducing localized hyperthermia, a powerful tool to produce heat. By applying an AMF to the anode electrode, the Pt-MNP material can harness this hyperthermia effect, allowing targeted electrochemical applications.

Our work focused on utilizing a novel catalyst comprising platinum-iron oxide magnetic nanoparticles supported on carbon cloth (CC) electrodes, with the added innovation of MagnetoElectroCatalysis assistance. This cutting-edge approach demonstrates a remarkable enhancement in the electrochemical conversion of ammonia, opening exciting possibilities for efficient and sustainable energy conversion technologies. Our findings contribute to the growing body of knowledge in electrocatalysis and potentially pave the way for future breakthroughs in this field.

## 2. Experimental Procedure

### 2.1. Pt, MNP, and Pt-MNP nanomaterials synthesis:

Platinum nanoparticles were simultaneously synthesized using the chemical reduction method with sodium borohydride [55]. The synthesis protocol was adapted from the procedure reported by Rodrigues et al. [49]. In this synthesis process, a solution of 12.5 mL of ethanol containing 0.2 mol L$^{-1}$ of H$_2$PtCl$_6$·6H$_2$O was added to a burette. Subsequently, 50 mL of ethanol was added to a round-bottom flask and placed in an ice bath at approximately 5 °C. After reaching thermal equilibrium, 10 mL of a cold aqueous NaBH$_4$ solution with a molar ratio of 5:1 (NaBH$_4$:Metal) was added to the flask. The system was then sealed with the burette, and N$_2$ gas was purged for 5 minutes to remove residual O$_2$ and stir the solution. Following this procedure, the burette was opened at a flow rate of 1 drop every 4 seconds. The product was centrifuged, washed twice with ethanol, and dried at 80 °C for 24 hours. The synthesis of iron oxide magnetic nanoparticles was carried out using the coprecipitation method [56]. Initially, 15 mL of a 0.53 mol L$^{-1}$ FeCl$_3$·6H$_2$O solution and 15 mL of a 0.27 mol L$^{-1}$ FeSO$_4$·7H$_2$O solution were combined in a 2:1 molar ratio and subjected to sonication for 30 minutes. Subsequently, 5 mL of 28% NH$_4$OH was gradually added at a rate of 2 drops per minute. The mixture was then sonicated for 10 minutes, centrifuged, and rinsed twice with a 1:4 (v/v) ethanol/water mixture and once with water. Finally, the resulting black product was dried at 80 °C for 24 hours. The preparation procedure for the CC working electrode began with the dispersion of 6 mg of electrocatalyst nanoparticles consisting of a physical mixture containing 80% MNP and 20% Pt m/m, which was dispersed in 345 µL of water, 145 µL of isopropanol, and 10 µL of a 5% Nafion® solution. This mixture was subjected to probe sonication. Subsequently, the dispersion was carefully applied to a 1 cm² area of the CC electrode in aliquots of 50 µL. After each aliquot deposition, the electrode was dried at 60°C for 10 minutes [57].

### 2.1. Physicochemical characterization

X-ray diffraction (XRD, Panalytical X'Pert Pro-PW3042/10) utilizing Cu Kα radiation with λ = 0.1540 nm set to 40 kV and 40 mA. Data acquisition spanned a 2θ range from 20 to 90°, with a scan rate of 0.02° per second. The morphology of the Pt-MNP catalyst was investigated by Scanning Electron Microscopy (SEM) using a JEOL JSM-7100F and Transmission Electron Microscopy (TEM) using a JEOL JEM 2100F electron microscope operating at 200 kV.

### 2.2. Electrochemical experiments

The experimental setup for electrochemical measurements involved utilizing an Autolab PGSTAT204 and a three-electrode electrochemical cell crafted from glass. The reference electrode was constructed as a Hg|HgO electrode, while a platinum foil was used as the counter electrode. An 8 cm long and 1 cm wide CC was employed for the working electrode responsible for electrocatalyst deposition. The electrochemical setup was integrated at the center of the hyperthermia coil, which could produce an AMF operating at a frequency of 224 kHz, with amplitudes ranging from 29.8 mT to 133.5 mT. These amplitudes were used to understand their influence on the electrocatalytic performance of Pt-MNP/CC in ammonia electro-oxidation. This investigation used cyclic voltammetry (CV) and chronoamperometry (CA) measurements. The CA experiments utilized Pt/CC, Pt-MNP/CC, and MNP/CC as the working electrodes. These tests were conducted over 2700 seconds, maintaining a potential of -0.30 V *vs* Hg|HgO, in the presence of a 35 mL solution containing 0.50 mol L$^{-1}$ ammonia in 1.0 mol L$^{-1}$ NaOH under the influence of an AFM. CV experiments were also conducted, which consisted of five cycles, with a scan rate of 0.010 V s$^{-1}$. The outcome of these electrocatalytic measurements was adjusted according to the activity catalyst's loading, as documented in existing literature about diverse reactions [58, 59].

## 3. Results and discussion
### 3.1. Physical characterization

Fig. 1 presents the XRD patterns of the synthesized Pt and MNP components. The Pt diffractogram exhibits all characteristic reflections of a face-centered cubic (FCC) structure at 39.8° (111), 46.3° (200), 67.5° (220), 81.3° (311), and 85.7° (222), in perfect agreement with the reference pattern for platinum (ICSD 241736). The MNP diffraction pattern shows peaks at 30.1° (220), 35.5° (311), 43.1° (400), 53.4° (422), 57.0° (511), 62.6° (440), and 74.0° (533), which match either maghemite ($\gamma$-$Fe_2O_3$, ICSD 172906) or magnetite ($Fe_3O_4$, ICSD 26410) phases. These iron oxide phases have nearly identical peak positions, so we are unable to suggest the predominant presence of either of them, in this way naming the compound as magnetic nanoparticles (MNP). The sharp, well-defined peaks in both patterns confirm the high crystallinity of the individual components prior to their physical combination to form the Pt-MNP composite. These results are consistent with previously reported studies [60-62] and validate the successful synthesis of both catalytic components.

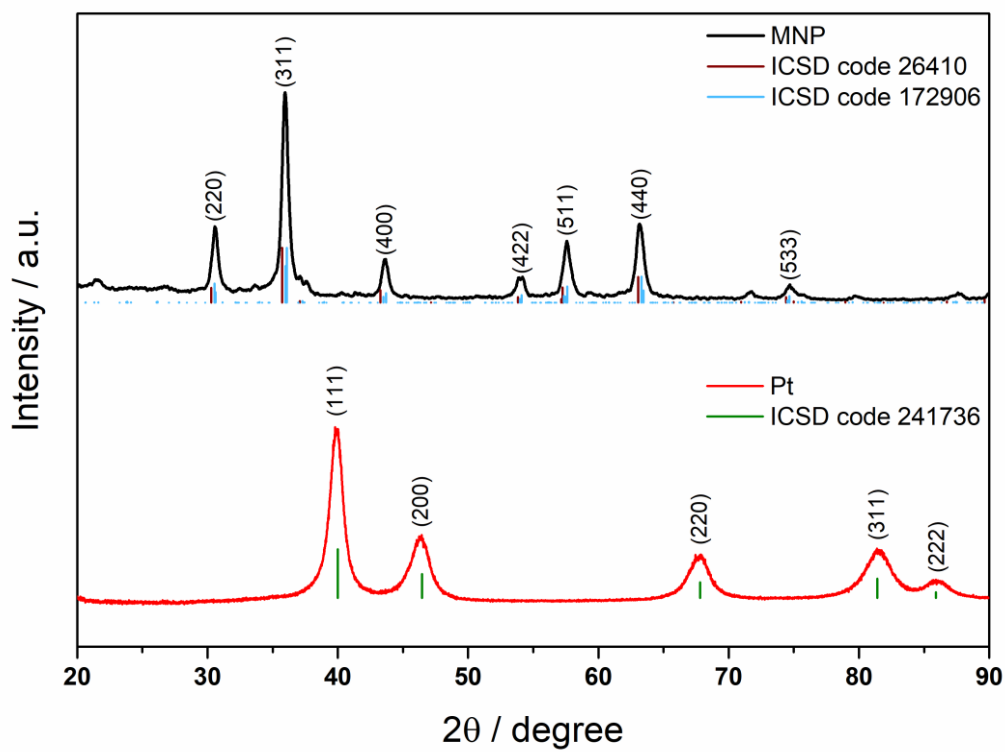

**Figure 1**. XRD patterns of magnetic nanoparticles and platinum nanoparticles.

The TEM (Fig. 2) effectively characterizes the Pt-MNP composite system. In Fig. 2a, MNP has an average size of about 21 nm, while the smaller Pt nanoparticles average around 8 nm. Fig. 2b and 2c confirm the structural integrity and distinct properties of the Pt NPs and MNPs. These results suggest that a straightforward physical mixture can yield a functional composite system with retained individual characteristics, providing a simple yet effective route for preparing hybrid nanomaterials.

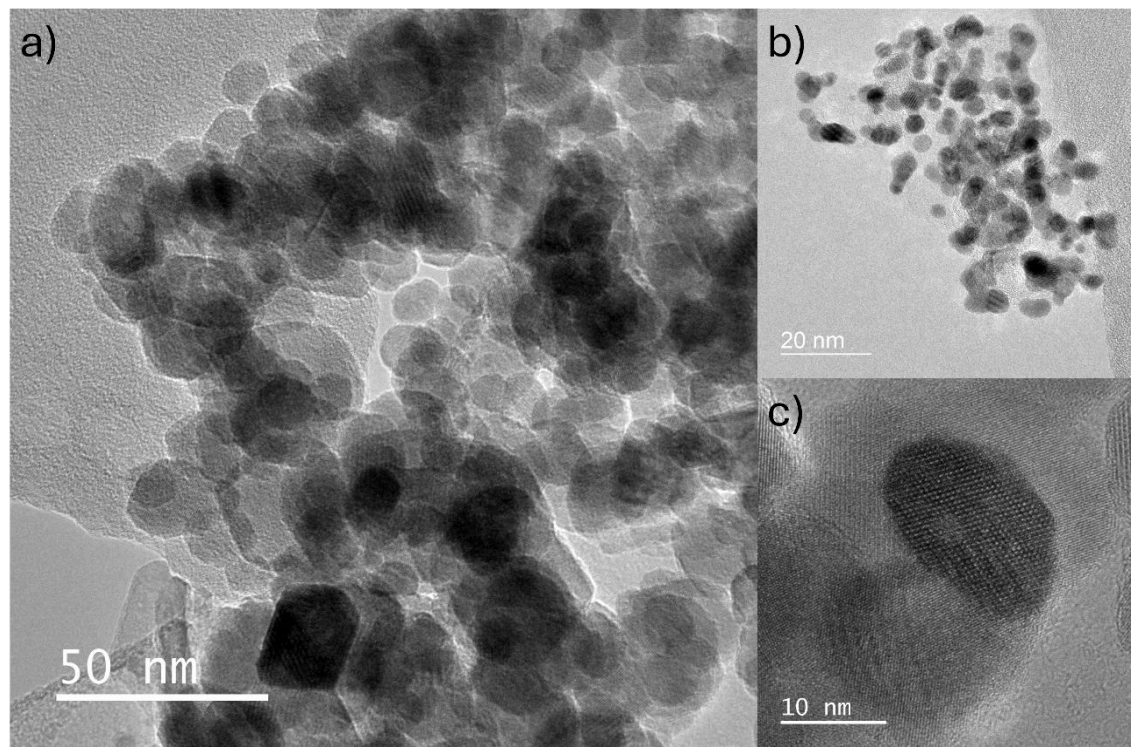

**Figure 2.** Low magnification TEM (a) of Pt-MNP and HRTEM (b and c) of Pt and MNP.

The intimate interaction between Pt nanoparticles and MNPs is further confirmed by STEM-EDS analysis (Fig. 3). The high-resolution STEM image (Fig. 3a) reveals Pt nanoparticles in close proximity to the larger MNPs, demonstrating their direct physical association. Elemental mapping (Figs. 3b–d) provides unambiguous evidence of the nanoscale spatial distribution, with Pt nanoparticles (Fig. 3b) precisely localized adjacent to the Fe- and O-rich MNP domains (Figs. 3c–d). This clear co-localization confirms that the Pt nanoparticles are not only in direct contact with the MNPs but are also distributed across their surfaces, reinforcing the formation of a well-integrated Pt-MNP composite system.

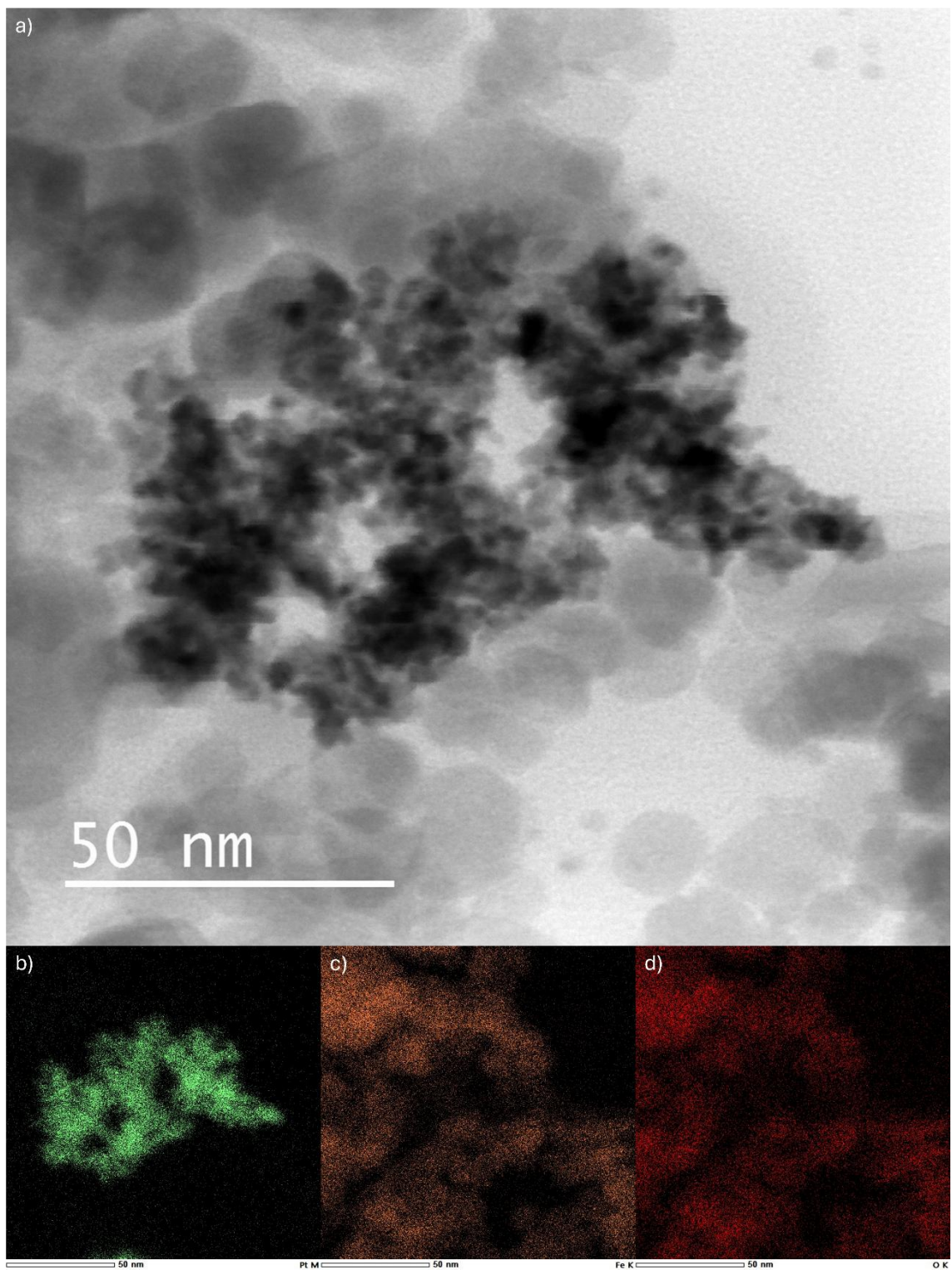

**Figure 3**. STEM-EDS analysis of the Pt-MNP composite: (a) STEM image showing Pt nanoparticles adjacent to MNPs. (b–d) EDS elemental maps for (b) Pt, (c) Fe, and (d) O, confirming the co-localization of Pt with the Fe- and O-rich MNPs.

The SEM images in Fig. 4 provide additional insight into the morphology and distribution of the Pt-MNP nanoparticles on the carbon cloth substrate. At lower magnification (x250, Fig. 4a), the fibrous structure of the carbon cloth is clearly visible, with Pt-MNP nanoparticles uniformly dispersed across its surface. Higher magnification images (x1000 and x3000, Figs. 4b and 4c) reveal the detailed morphology of the nanoparticles, confirming their successful deposition and distribution. The varying magnifications highlight the interconnected network of the carbon cloth fibers and the effective anchoring of the Pt-MNP composite, which is critical for enhancing surface area and catalytic performance.

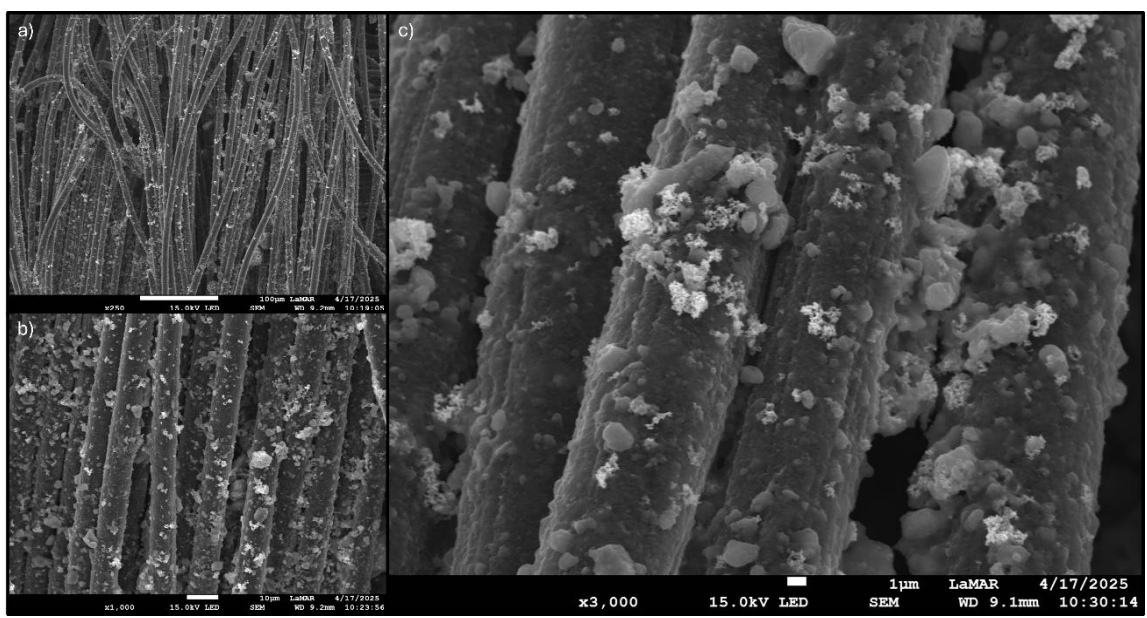

**Figure 4.** SEM images of Pt-MNP nanoparticles dispersed on carbon cloth at different magnifications: (a) Low-magnification (x250), (b) Intermediate magnification (x1000), and (c) High-magnification (x3000). Scale bars are included for reference.

### 3.2. Electrochemistry

To electrochemically characterize the Pt and Pt-MNP after the deposition on CC, cyclic voltammetry experiments were performed in 1 mol L$^{-1}$ NaOH in a potential window of -0.85 V to 0.20 V, as shown in Fig. 5.

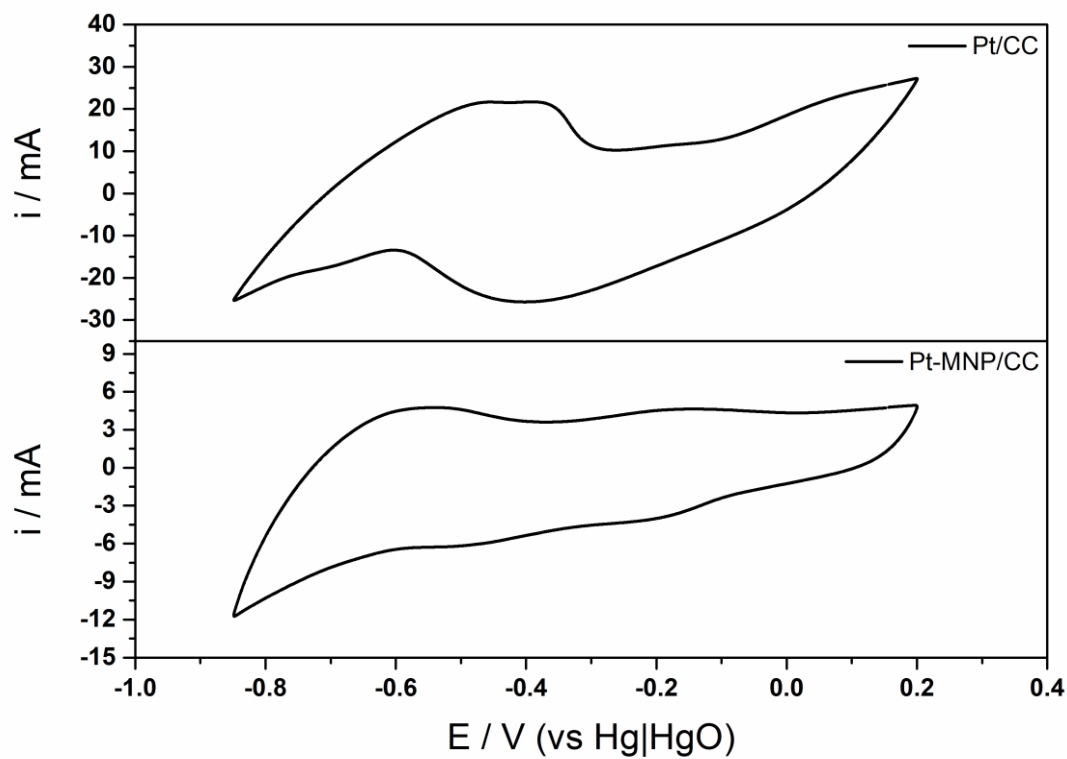

**Figure 5**. Cyclic voltammograms of Pt/CC and Pt-MNP/CC in 1 mol L$^{-1}$ NaOH at ν =50 mV s$^{-1}$.

Fig. 5 revealed intriguing insights into the electrochemical behavior of the catalyst materials. For both Pt/CC and Pt-MNP/CC, the two peaks associated with the hydrogen adsorption/desorption process are not well defined. The absence of well-defined peaks can be attributed to the direct deposition of nanoparticles onto the carbon cloth electrode, which results in a less ordered surface structure compared to traditional smooth electrodes. In the case of Pt/CC, the lack of distinct peaks suggests that the platinum nanoparticles are dispersed in a manner that limits the formation of well-defined crystalline facets, which are typically responsible for sharp hydrogen adsorption/desorption features. For Pt-MNP/CC, the interaction between platinum and iron oxide at the electrode surface further modulates the voltametric response, potentially altering the electronic properties of platinum and contributing to the broadening of the peaks. This behavior is consistent with previous reports for platinum nanoparticles supported on carbon cloth [63], as well as platinum combined with various oxides [64, 65].

Fig. 6 shows the cyclic voltammograms in 1 mol L$^{-1}$ NaOH with 0.5 mol L$^{-1}$ NH4OH. The results include the experiment in the absence and presence of different AMFs (74.0, 88.9, 103.9, 118.4, and 133.5 mT).

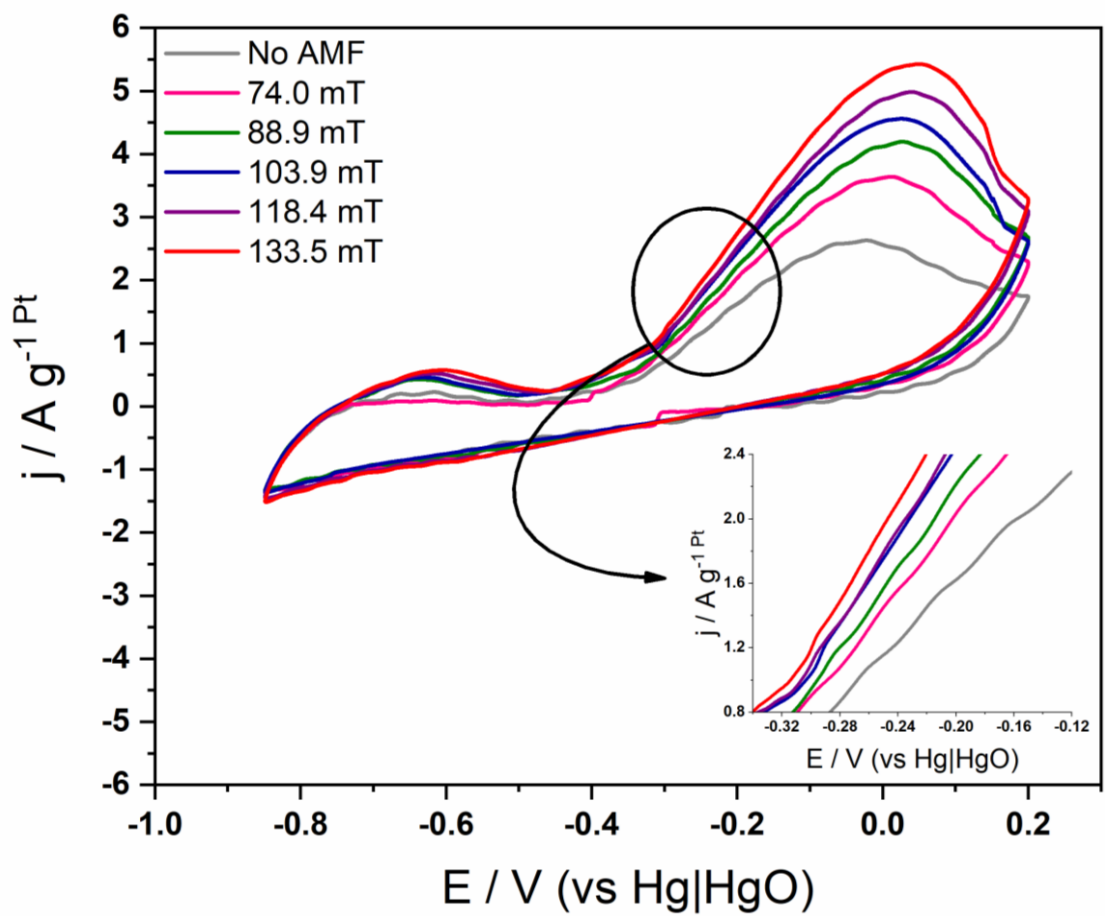

**Figure 6**. Cyclic voltammograms of Pt-MNP/CC in 1 mol L$^{-1}$ NaOH with 0.5 mol L$^{-1}$ NH$_4$OH in the absence and in the presence of different AMFs (74.0, 88.9, 103.9, 118.4, and 133.5 mT) at v = 20 mV s$^{-1}$ with a frequency of 224 kHz.

The application of AMFs in electrochemical systems has attracted significant research interest due to their ability to enhance catalytic performance. As shown in Fig. 6, the Pt-MNP/CC catalyst exhibits a clear trend in which increasing AMF intensity correlates with a higher peak current density for the AEOR. This result suggests that the AMFs enhance reaction kinetics and introduce a means to tune electrocatalytic activity dynamically. Notably, the observed increase in peak current density under higher magnetic field amplitudes (74.0, 88.9, 103.9, 118.4, and 133.5 mT) highlights a direct relationship between magnetic field intensity and AEOR efficiency.

The results demonstrate a systematic enhancement in peak current density with increasing AMF strength. The peak current was measured at 2.62 A g$^{-1}$ Pt without an alternate magnetic field. With the application of 74 mT, the peak current increased to 3.63 A g$^{-1\ Pt}$, representing a 38.5% gain. For 88.9 mT, the peak current further increased to 4.19 A g$^{-1\ Pt}$, a 59.9% improvement compared to the baseline. At 103.9 mT, the value rose to 4.56 A g$^{-1\ Pt}$, equating to a 74.0% increase. Applying 118.4 mT yielded 4.98 A g$^{-1\ Pt}$, an impressive 90.1% enhancement, while the highest tested field of 133.5 mT produced a peak current of 5.43 A g$^{-1\ Pt}$, more than doubling the initial value with a 107.3% increase. These results underline the significant influence of AMFs in boosting the catalytic performance of the system.

The influence of AMFs on peak current density reinforces their role as a tool for precision control and optimization in electrocatalytic reactions, positioning AMFs as a promising way for modulating reaction kinetics in catalytic systems [66]. The dynamic interaction between magnetic fields and electrochemical reactions introduces novel prospects for developing magnetically responsive electrocatalytic systems, where catalytic activity can be modulated in situ based on magnetic field strength.

The inserted graph in Fig. 6 illustrates the AMF-induced shifts in the onset potential for AEOR, demonstrating that higher AMF amplitudes drive the onset potential toward more negative values. This shift implies a reduced overpotential, thereby improving the thermodynamics of ammonia electro-oxidation [67,68]. The trend of increasingly negative onset potentials with higher AMF amplitudes reflects an enhanced driving force for AEOR, further indicating that AMFs can effectively modulate the thermodynamic aspect of the reaction [69,70].

The importance of localized heating provided by the AMF emerges as a crucial factor influencing the observed onset potential shifts in the electro-oxidation of ammonia. The application of AMFs induces localized heating at the catalyst-electrolyte interface, which can significantly impact the energetics of the electrochemical reaction [49]. Furthermore, the degree of localized heating is inherently linked to the intensity of the applied magnetic field, amplifying the magnitude of the observed potential shifts with increasing AMF strength. Thus, the combined effects of magnetic field-induced modulation of reaction kinetics and localized heating underscore the multifaceted influence of AMFs on the electrocatalytic performance [71].

The cyclic stability of the Pt-MNP/CC catalyst under MagnetoElectroCatalytic conditions was rigorously evaluated through ten consecutive CV cycles at the optimal AMF intensity (Fig. S1). The near-perfect overlap of successive cycles demonstrates exceptional operational stability, evidenced by three key performance metrics: (1) extremely low variation in electroactive surface area, (2) stable onset potentials for ammonia oxidation, and (3) consistent current densities (±1.6%) across all cycles. This remarkable stability directly correlates with the catalyst's sustained performance, confirming that the magnetically-enhanced interface between Pt and MNPs maintains its structural integrity and catalytic activity. The absence of peak shifts or current decay further verifies that neither Pt poisoning nor iron oxide dissolution occurs under operating conditions, making this system particularly promising for long-term applications.

The initial chronoamperometry investigation was conducted under specific conditions: 1 mol L$^{-1}$ NaOH solution containing 0.5 mol L$^{-1}$ NH$_4$OH, under a potential of - 0.2 V for 4800 seconds. As depicted in Fig. 7 the influence of AMFs with different amplitudes (29.8, 44.8, 56.6, 74.0, 88.9, 103.9, 118.4, and 133.5 mT) was investigated. Each magnetic field was activated for 300 seconds before being systematically deactivated, this cycle meticulously repeated eight times across the range of AMF amplitudes.

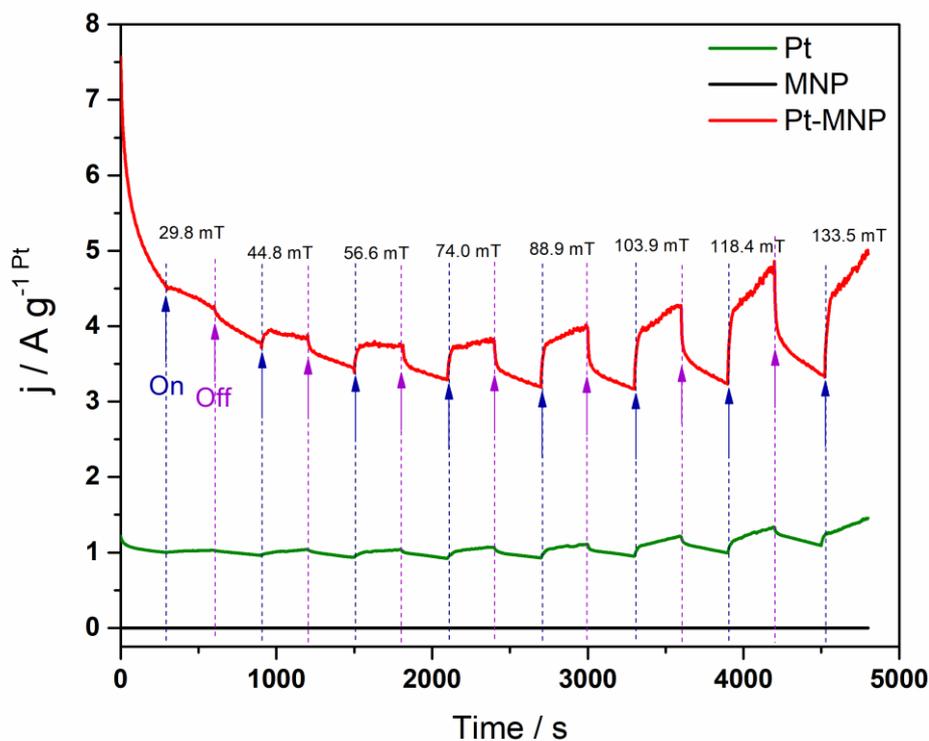

**Figure 7**. Chronoamperometry measurement of Pt-MNP/CC in 1 mol L$^{-1}$ NaOH with 0.5 mol L$^{-1}$ NH$_4$OH carried out at -0.2 V in the absence and in the presence of different AMFs (29.8, 44.8, 56.6, 74.0, 88.9, 103.9, 118.4, and 133.5 mT) with a frequency of 224 kHz.

As illustrated in Fig. 7, a higher magnetic field value correlates directly with a proportionally more significant increase in the resulting current. Notably, the initial stage of the process (approximately 40 seconds after activation of the AMF) exhibits a pronounced surge, followed by a subsequent phase characterized by a distinct angular coefficient. In this scenario, the abrupt enhancement of the obtained current within the first 40 seconds of activating the AMF can be attributed to the interplay between the Néel relaxation mechanism, which generates heat in iron oxide magnetic nanoparticles, and the dynamics of ammonia electro-oxidation on the platinum nanoparticles deposited on the carbon cloth electrode [72].

Initially, when the AMF is switched on, it induces rapid oscillations of the magnetic moments associated with the magnetic nanoparticles. As the MNPs are immobilized on the CC, we can expect negligible heat generation due to the Brown mechanism. Otherwise, the Néel relaxation creates highly localized heating only within the MNPs as the magnetic moments attempt to realign with the changing magnetic field. This localized heating is highly efficient, so that within the first few seconds of AMF activation, the MNP can attain temperatures as high as hundreds of degrees [73]. Consequently, the surrounding ammonia molecules undergo rapid oxidation on the surface of the platinum nanoparticles. Thus, the elevated temperature enhances the AEOR, enabling it to achieve higher current densities from the related process.

During this initial period, the availability of ammonia molecules near the electrode surface is high, and the rate of ammonia electro-oxidation is primarily limited by the kinetics of the surface reaction. Therefore, the rapid oxidation of ammonia driven by the AMF leads to a sudden enhancement in the obtained current within the first 40 seconds. However, as AEOR progresses and the concentration of ammonia molecules near the electrode surface decreases due to its consumption, the diffusion of ammonia from the bulk solution to the electrode becomes increasingly rate-limiting. This diffusion-controlled regime becomes more dominant after the initial period of rapid surface reaction kinetics [74]. Consequently, beyond the initial 40 seconds, the enhancement of the obtained current becomes less pronounced. The decrease in the angular coefficient of the current enhancement can be attributed to the transition from a primarily surface reaction-limited regime to a diffusion-limited regime as the electro-oxidation reaction proceeds [75].

In summary, the abrupt current enhancement observed within the first 40 seconds of AMF activation demonstrates Néel relaxation-driven localized heating as the dominant mechanism [76-78], where thermal energy from MNP rapidly activates adjacent Pt sites (Fig. 7). This creates a transient kinetic regime where surface reaction rates are enhanced through: (i) thermal activation of $NH_3$ dehydrogenation steps, and (ii) increased mobility of adsorbed intermediates. The subsequent current stabilization reflects the transition to diffusion-limited operation as near-surface ammonia becomes depleted - a characteristic signature of thermally-enhanced electrocatalysis [79,80].

The control experiments with Pt/CC reveal fundamentally different behavior (Fig. 7). While minimal current enhancement occurs at lower field, slight activity increases emerge at our highest applied fields. This can be attributed to platinum's inherent magnetic properties. Though weakly paramagnetic, Pt nanoparticles exhibit field-enhanced spin polarization at high AMF intensities, potentially modifying d-band center positions. However, these effects are orders of magnitude smaller than in Pt-MNP/CC.

These results establish that AMF enhancement in Pt-MNP/CC stems primarily from nanoscale thermal coupling between magnetically heated MNP and catalytic Pt sites, with minimal contribution from pure field effects. The kinetic signatures confirm that the Néel relaxation mechanism provides both the spatial and temporal resolution needed to dynamically modulate electrocatalytic processes at millisecond timescales.

To further elucidate the aforementioned points, another chronoamperometry experiment was performed, wherein a fixed AMF amplitude of 103.9 mT was applied. The experimental protocol involved an initial period of 300 seconds without the application of the AFM, followed by a 1200-second under the AMF, and concluded with a subsequent 300-second period with the magnetic field turned off, as depicted in Fig. 9.

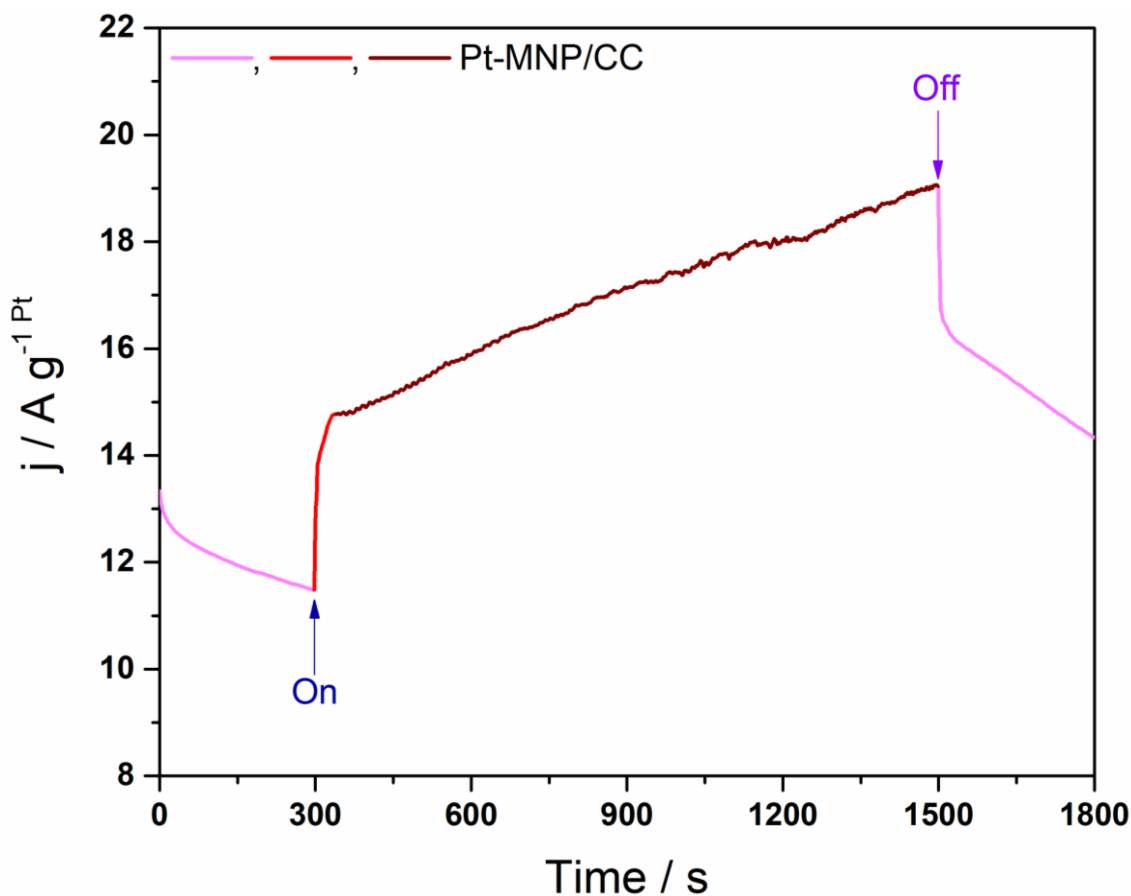

**Figure 8**. Chronoamperometry measurement of Pt-MNP/CC in 1 mol L$^{-1}$ NaOH with 0.5 mol L$^{-1}$ NH$_4$OH was carried out at -0.2 V in the absence and presence of an AMF of 103.9 mT with a frequency of 224 kHz.

In this latest chronoamperometry result, a distinct pattern emerges, reaffirming the observations previously discussed regarding the interplay between the Néel relaxation mechanism and the dynamics of ammonia electro-oxidation. Once again, upon activation of the AMF, there is a notable surge in the obtained current within the initial 40 seconds (red line in Fig. 8), elevating from 11.5 to 14.8 A g$^{-1}$ $^{Pt}$, a 1.3-fold increase. This abrupt increase underscores the significant role of the localized heating of iron oxide magnetic nanoparticles under AMF, facilitating the rapid oxidation of adjacent ammonia molecules on the platinum nanoparticles. Subsequently, over 1160 seconds of the experiment (wine line in Fig. 8), a more gradual increase in the obtained current is observed, rising from 15 to 19.0 A g$^{-1}$ $^{Pt}$, a 1.25-fold increase. This slower rate of increase in current further supports the notion that diffusion processes become increasingly influential as the electro-oxidation reaction progresses, emphasizing the complex interplay between surface reaction kinetics and bulk diffusion dynamics in the ammonia electro-oxidation system under the influence of an AMF [81].

While previous studies [82-85], have demonstrated that global heating of the electrochemical system can enhance performance by accelerating reaction kinetics, such methods typically require prolonged heating durations (on the order of hours) to achieve measurable improvements in current density. In contrast, the MagnetoElectroCatalytic approach for AEOR presented here leverages the Néel relaxation mechanism to induce rapid, localized heating of iron oxide nanoparticles under an AMF, enabling a near-instantaneous (within seconds) current density enhancement. This localized thermal effect not only circumvents the energy inefficiency associated with bulk solution heating but also minimizes unwanted side reactions or degradation that may arise from prolonged thermal exposure [49, 50].

## 4. Conclusions

This study demonstrates that platinum-iron oxide nanoparticles (Pt-MNP) under an alternating magnetic field (AMF) significantly enhance ammonia electro-oxidation through a synergistic MagnetoElectroCatalysis mechanism. The AMF generates localized heat via Néel relaxation, boosting reaction kinetics and catalytic efficiency, as confirmed by cyclic voltammetry and chronoamperometry. Compared to conventional methods, this MagnetoElectroCatalysis approach offers superior performance and stability for ammonia removal. Our findings highlight the potential of magnetic-field-assisted electrocatalysis for environmental remediation and sustainable energy applications.

**Supplementary Material:** Supplementary data for this article can be found online.

**CRediT authorship contribution statement**

**Caio Machado Fernandes:** Conceptualization, Methodology, Formal analysis, Investigation, Validation, Data curation, Visualization, Funding acquisition, Writing – original draft, Writing – review & editing. **Eduardo M. Rodrigues:** Methodology, Formal analysis, Investigation, Validation. **Odivaldo C. Alves:** Conceptualization,

Methodology, Formal analysis, Writing – Original Draft, Visualization, Writing – review & editing. **Flavio Garcia:** Conceptualization, Supervision, Resources, Funding acquisition, Writing – review & editing. **Yutao Xing:** Investigation, Validation, Resources, Writing – review & editing. **Mauro C. Santos:** Funding acquisition, Supervision, Writing – review & editing. **Julio Cesar M. Silva:** Conceptualization, Methodology, Resources, Supervision, Funding acquisition, Writing – review & editing, Supervision.

## Acknowledgments

The authors would like to thank Fundação de Amparo à Pesquisa do Estado de São Paulo (FAPESP, #2022/10484-4, #2024/03549-8, #2022/15252-4, #2022/12895-1) and Conselho Nacional de Desenvolvimento Científico e Tecnológico (CNPq, #308663/2023–3, #402609/2023–9) for the financial support.